\documentclass{nature}

\usepackage{amsmath}    
\usepackage{graphicx}   
\usepackage{color}
\usepackage[10pt]{moresize}
\usepackage{graphics,graphicx}
\usepackage{amsfonts}
\usepackage{amssymb}
\usepackage{amscd}
\usepackage{amsmath}
\usepackage{enumerate}
\usepackage{epsfig}
\usepackage{subfigure}

\let\oldcaption\caption
\renewcommand{\caption}{\sffamily \oldcaption}
\begin{document}

\title{Phase self-aligned continuous-variable measurement-device-independent quantum key distribution}

\author{Hua-Lei Yin, Wei Zhu \& Yao Fu}

\maketitle

\begin{affiliations}
\item
Department of Physics, Zhejiang Institute of Modern Physics and ZJU-Phoenix Synergetic Innovation Center in Quantum Technology, Zhejiang University, Hangzhou 310027, China\\
Correspondence and requests for materials should be addressed to H.-L.Y. (email: hlyin@zju.edu.cn) or Y. F. (email: yaofu@mail.ustc.edu.cn)
\end{affiliations}

\baselineskip24pt

\maketitle

\begin{abstract}
Continuous-variable measurement-independent-device quantum key distribution (CV-MDI-QKD) can offer high secure key rate at metropolitan distance and remove all side channel loopholes of detection as well. However, there is no complete experimental demonstration of CV-MDI-QKD due to the remote distance phase-locking techniques challenge. Here, we present a new optical scheme to overcome this difficulty and also removes the requirement of two identical independent lasers.
Furthermore, we give an alternate but detailed proof of the minimized key rate condition to extract the secure key rate.
We anticipate that our new scheme can be used to demonstrate the in-field CV-MDI-QKD experiment and build the CV-MDI-QKD network with untrusted source.
\end{abstract}

Quantum key distribution (QKD) allows remote parties (usually referred as Alice and Bob) to establish a string of secure key even at the presence of an eavesdropper (referred as Eve)~\cite{Gisin:2002:Quantum,Scarani:2009:The,Weedbrook:2012:Gaussian}.
Based on the laws of quantum mechanics, QKD achieves formidable task and opens a whole new area which is under rapid development and attracting lots of attentions~\cite{qiu:2014:quantum}. Together with the well-known one time pad cryptosystem~\cite{shannon:1949:communication}, QKD may provide information-theoretic security. In reality, the side-channel of the QKD system are the major security risks~\cite{Fung:2007:Phase,Zhao:2008:Quantum,lydersen:2010:hacking,weier:2011:quantum,Tang:2013:Source}. Fortunately, a novel protocol named measurement-device-independent QKD (MDI-QKD) ~\cite{Braunstein:2012:Side,Lo:2012:Measurement,Zhou:2016:Making,lo:2014:secure} has been presented to remove all side channel attacks of detectors. In the MDI-QKD protocol, both Alice and Bob prepare perfect quantum states and then send them to the untrusted relay (referred as Charlie) through insecure quantum channels for Bell state measurement (BSM), the correlation between Alice and Bob can be built through the BSM results announced by Charlie. Together with the decoy-state theory~\cite{Wang:2005:Beating,Lo:2005:Decoy}, the qubit-based discrete-variable (DV) MDI-QKD has been successfully experimentally demonstrated~\cite{Rubenok:2013:Real,Liu:2013:Experimental,Ferreira:2013:Proof,Tang:2014:Experimental,Tang:2014:Measurement,Wang:2015:Phase,Yin:2016:Measurement, comandar:2016:quantum}. Meanwhile, MDI-QKD also opens a new road to help building quantum communication network \cite{Fu:2015:Long,Tang:2016:Measurement,Yin:2017:Experimental,roberts:2017:experimental}, even though, in the basic MDI-QKD protocols~\cite{Braunstein:2012:Side,Lo:2012:Measurement}, the untrusted relay is a passive repeater which is not able to beat the point-to-point repeater-less bound \cite{pirandola2017fundamental} (see also a review \cite{pirandola2018theory}).

The experimental demonstration of DV-MDI-QKD has been successfully performed over 404 km optical fiber~\cite{Yin:2016:Measurement}, while the secure key rate is relatively low even at metropolitan distance due to the single photon encoding. Continuous variable (CV) encoding scheme can also be used for QKD to share secret key~\cite{Cerf:2001:Quantum,Grosshans:2002:Continuous,grosshans:2003:quantum,Grosshans:2004:Continuous,Weedbrook:2004:Quantum,pirandola:2008:continuous,Renner:2009:de,Jouguet:2011:Long,Leverrier:2015:Composable,Qi:2015:Generating,Soh:2015:Self}. The Gaussian-modulated coherent states CV-QKD has been successfully experimentally demonstrated through approximate 100 km optical fiber~\cite{jouguet:2013:experimental,huang:2016:long} under collective attack.
Recently, the security of coherent states continuous-variable quantum key distribution against coherent attack in a realistic finite-size regime has been proved~\cite{Leverrier:2017:Security},
which paves the way to information-theoretically secure CV-QKD. Due to its obvious advantages of high secure key rate, low cost and running in room-temperature, CV-QKD becomes a very active research area in quantum information~\cite{Weedbrook:2012:Gaussian}.
Similar with the DV system, several practical attacks about detectors are also found in CV-QKD recently~\cite{Jouguet:2013:Preventing,Ma:2013:Local,Huang:2014:Quantum,Qin:2016:Quantum}. In order to be immune to all detector loopholes and achieve high secure key rate at metropolitan distance, the
CV-MDI-QKD is proposed recently~\cite{pirandola:2015:high,Ma:2014:Gaussian,Li:2014:Continuous,Ottaviani:2015:Continuous} with a proof-of-principle experimental demonstration performed on the free space optical platform by sharing a highly stable laser~\cite{pirandola:2015:high}. The CV-MDI-QKD has shown great potential for its high secret key rate and low commercial cost compared with the corresponding DV-MDI-QKD protocol~\cite{pirandola:2015:high,xu:2015:discrete,pirandola:2015:reply}.
However, a complete experimental demonstration of CV-MDI-QKD with distances of kilometers in optical fiber or free space is still not yet realized due to technical challenges.
Thereinto, the main challenge is that remote distance phase-locking technique is necessary to achieve correct CV BSM. However, this requirement is not needed in DV-MDI-QKD due to the fact that the two-photon Hong-Ou-Mandel interference is phase unrelated~\cite{Zhao:2007:Robust}.

In this paper, we present a new optical scheme to solve the common phase-reference problem. The central idea of this new optical scheme is that reliable phase reference can be established by transmitting the two laser pulses quantum states through the same optical path, such as optical fiber or free space. Compared with the previous protocol~\cite{pirandola:2015:high}, a pair of Faraday mirrors (FM) and several polarization beam splitters (PBS) are needed~\cite{Muller:1997:plug,Choi:2016:plug}. No complicated active or passive phase-locking or compensation technology is required in this new scheme, which greatly simplifies the experimental realization of CV-MDI-QKD. Similar structures has already been successfully used in a proof-of-principle experimental demonstration of quantum fingerprinting beating the classical limit~\cite{Guan:2016:Observation}. Furthermore, we give an alternate but detailed proof of the minimized key rate condition to extract the secure key rate, which has been used and first proved in Ref. 49.

\section*{Results}
\begin{figure*}
\centering
\resizebox{12cm}{!}{\includegraphics{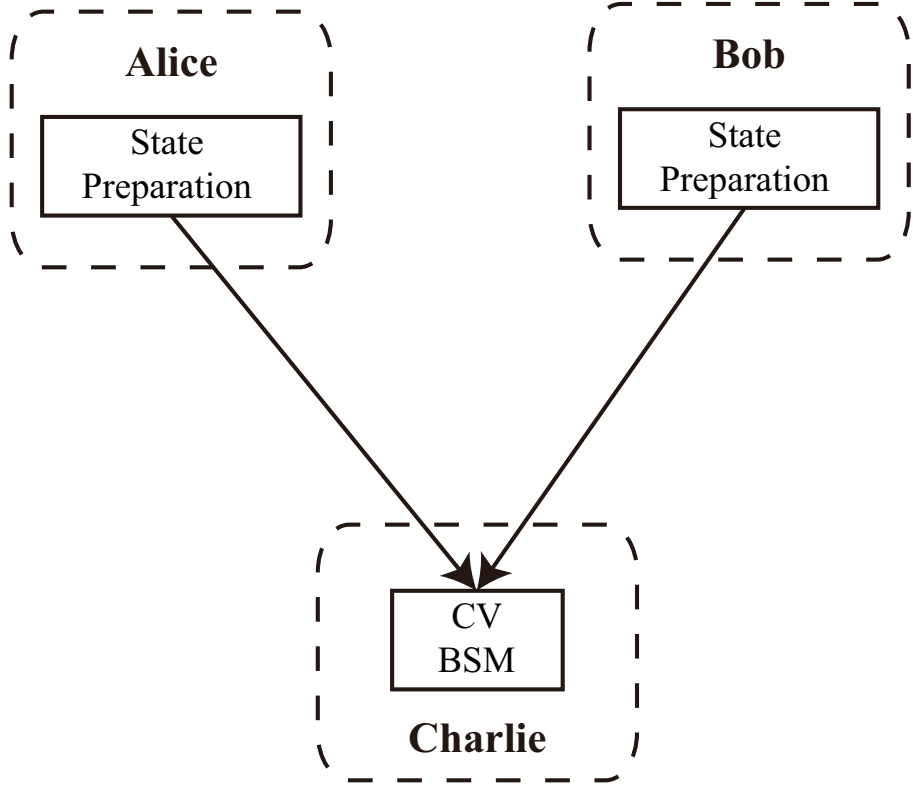}}
\caption{The general structure of the CV-MDI-QKD protocol.}
\label{fig:1}
\end{figure*}

\textbf{Phase self-aligned CV-MDI-QKD.} The basic idea of CV-MDI-QKD protocol is shown in Fig.~\ref{fig:1}. It includes the following major steps~\cite{pirandola:2015:high,Ma:2014:Gaussian,Li:2014:Continuous}: (1) Alice and Bob  prepare the sequence of Gaussian-modulated coherent states $|\hat{x}_A+i\hat{p}_A\rangle$ and $|\hat{x}_B+i\hat{p}_B\rangle$, respectively. (2) The quantum states are transmitted through insecure quantum links to the untrusted relay for CV BSM. (3) The incoming quantum states from two parties are coupled to interfere through a balanced beam splitter. Two output quantum states are measured with homodyne in $\hat{x}$ and $\hat{p}$ quadrature, with $\hat{x}_{-}=(\hat{x}_A-\hat{x}_B)/\sqrt{2}$ and $\hat{p}_{+}=(\hat{p}_A+\hat{p}_B)/\sqrt{2}$. The measurements together give $\gamma=(\hat{x}_{-}+i\hat{p}_{+})/\sqrt{2}$ with the probability $p(\gamma)$, and the results $\gamma$ are transmitted to both Alice and Bob through the public channel. (4) Based on the measurement results, Alice or Bob confirms the other's state value, then correlation is built. (5) Similar with the usual CV-QKD protocol, through error correction and privacy amplification, a secure key is acquired.

\begin{figure*}
\centering
\resizebox{12cm}{!}{\includegraphics{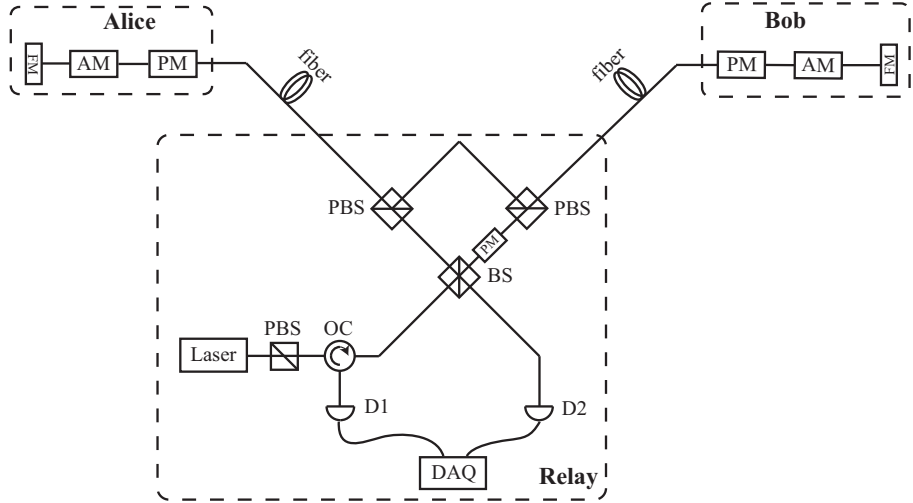}}
\caption{Schematic structure of the self-aligned CV-MDI-QKD protocol. PBS, polarization beam splitter; OC, optical circulator; BS, 50:50 beam splitter;
AM, amplitude modulator; PM, phase modulator; FM, Farady mirror; D1, D2, detector; DAQ, data acquisition.}
\label{fig:2}
\end{figure*}

Fig.~\ref{fig:2} shows the detailed structure of our new scheme with optical fiber. The laser pulse goes through a polarization maintaining fiber PBS (PMF-PBS) to filter out its horizonal (H) polarized component. Then after a polarization maintaining optical circulator, the laser pulse directly goes through a balanced PMF beam splitter (PMF-BS) and be splitted into two identical laser pulses that each will be finally transmitted to Alice and Bob for CV encoding. Meanwhile, this PMF-BS will be reused in the CV BSM later. For simplicity, we name the two identical laser pulses as the left and right pulse.
Here, we take the optical transmission process of the left pulse as an example. The left pulse goes out of the PMF-BS and into a PMF-PBS, due to its H polarization, it transmits through the PMF-PBS.
Through a long distance single mode optical fiber, the left pulse goes into Alice's port. All the modulators are not working at this moment. The left pulse is then reflected by the FM, changing its polarization to vertical (V) polarization. Once the left pulse is transmitted back to the PMF-PBS, due to the change of its polarization it is reflected this time and going to the right branch of this set. Then it meets another PMF-PBS and again reflected. The left pulse is transmitted to Bob's port through another long distance single mode optical fiber, then it is reflected by FM. Once reflected, the polarization is restored back to H, and Bob prepares his CV Gaussian-modulated quantum states. The modulated left pulse goes back to the PMF-PBS, due to its H polarization, it transmits through the PMF-PBS and back to the balanced PMF-BS of untrusted relay for CV BSM. As for the right pulse, the process is quite similar, except that the modulation process takes place at Alice's port. The CV quantum states prepared by Alice and Bob will stably interfere due to the two identical laser pulses go through the same optical path. In another word, the phase-reference is self-aligned. By the way, a $\pi/2$ phase modulation to the encoded pulse back from Bob's port is added for CV Bell detection purpose~\cite{pirandola:2015:high}. For simplified, the additional $\pi/2$ phase can be directly implemented in Bob's quantum states preparation stage. It's worth noting that the idea of our new optical scheme about self-aligned phase-reference can also be used in free space and satellite-based CV-MDI-QKD~\cite{Hosseinidehaj:2017:CVMDI,bedington:2017:progress,liao:2017:satellite,Liao:2018:Satellite}, especially for geostationary satellite~\cite{gunthner2017quantum}.

\noindent
\textbf{The secure key rate and simulation result.}
The security of general CV-MDI-QKD protocol has been analyzed under two kinds of quantum attack. One is using two independent entanglement cloners~\cite{Ma:2014:Gaussian,Li:2014:Continuous}, each attacking the channel between the relay and Alice or Bob independently. This kind of attack just extends the one-channel entanglement cloner attack in CV-QKD protocol to two channels independently. The other kind of attack is more general~\cite{pirandola:2015:high}, two ancillary modes from a reservoir of ancillas each attacks one of the two channels. These two ancillary modes may be correlated, which means the first kind of attack is nothing but a special condition of the second kind of attack. It has been proven that the most effective attack is when the two ancillary modes are entangled~\cite{pirandola:2015:high}, and this is also the most powerful attack of Eve can launch in the CV-MDI-QKD protocol.

Here, we follow the security analysis in Ref. 49. For general two channel attack, the two Gaussian ancillas can be described by the following covariance matrix~\cite{pirandola:2015:high}
\begin{equation}
\begin{aligned}
\sigma_{E_{1}E_{2}}=\left(
                  \begin{array}{cc}
                    \omega_A\textbf{I} & \textbf{G} \\
                    \textbf{G} & \omega_B\textbf{I} \\
                  \end{array}
                \right),
\end{aligned}
\end{equation}
with $\textbf{I}=\left(\begin{array}{cc}1 & 0 \\0 & 1 \\\end{array}\right)$ and $\textbf{G}=\left(\begin{array}{cc}g & 0 \\0 & g' \\\end{array}\right)$.
Where $\omega_{A}$ and $\omega_{B}$ are the eigen-frequencies of these two ancillas, while $G$ stands for the correlation between two ancillary modes with corresponding parameters $g$ and $g'$.

Here we assume that Alice's raw key is the reference. For simplify, we ignore the Trojan-horse attack in the quantum states preparation. The Trojan-horse attack of untrusted source in plug-and-play structure will be discussed in discussion section later.
With the ideal Gaussian modulation variance $\varphi \gg 1$, the secure key rate of CV-MDI-QKD protocol $R$ can be given by~\cite{pirandola:2015:high}
\begin{equation}
\begin{aligned}
R&=\xi I_{AB}-I_{EA},\\
I_{AB}&=\log_2\frac{\mu}{\chi},\\
I_{EA}&=h\left(\frac{\sqrt{\lambda\lambda'}}{|\tau_A-\tau_B|}\right)+\log_2\frac{e|\tau_A-\tau_B|\mu}{2(\tau_A+\tau_B)}-h(\nu),
\end{aligned}
\end{equation}
with $I_{AB}$ representing the mutual information between Alice and Bob, while $I_{EA}$ being Eve's information about the raw key of Alice.
$\xi\leq 1$ stands for the reconciliation efficiency,
$\chi$ is the equivalent noise, $\tau_A$ and $\tau_B$ are Alice's and Bob's transmissivities, respectively. The $h$-function is defined as $h(x)=\frac{x+1}{2}\log_2\frac{x+1}{2}-\frac{x-1}{2}\log_2\frac{x-1}{2}$ and $\mu=\varphi+1$. Some intermediate
variables are introduced to simplify the above equations, their definitions are~\cite{pirandola:2015:high}
$\nu=\sqrt{(\tau_A+\lambda)(\tau_A+\lambda')}/\tau_B$, $\lambda=\kappa-ug$, and $\lambda'=\kappa+ug'$,
with $\kappa=(1-\tau_A)\omega_A+(1-\tau_B)\omega_B$, and $u=2\sqrt{(1-\tau_A)(1-\tau_B)}$.
In the proof-of-principle CV-MDI-QKD experiment~\cite{pirandola:2015:high}, the modulation variance $\varphi=60$ has been achieved, so the ideal modulation assumption is a good approximation.
As for reconciliation efficiency $\xi=0.97$ can be achieved~\cite{Jouguet:2011:Long}, which is very close to $1$. The difference seems minor but has a major impact on the rate, improving $\xi$ is a major task in nowadays CV-QKD research.
When the transmission efficiency $\tau_{A}=\tau_{B}=\tau$, i.e., the symmetric condition, the secure key rate of CV-MDI-QKD under the realistic condition can be given by
\begin{equation}
\begin{aligned}
R=\log_2\frac{8\tau\mu^{\xi-1}}{e^2\chi^{\xi}\sqrt{\lambda\lambda'}}+h(\nu_1)\label{aa3},
\end{aligned}
\end{equation}
with $\nu_{1}=\sqrt{(\tau+\lambda )(\tau+\lambda ^{\prime})}/\tau$. When the transmission efficiency $\tau_{A}\neq\tau_{B}$, i.e., the asymmetric condition, the secure key rate of CV-MDI-QKD under the realistic condition can be given by
\begin{equation}
\begin{aligned}
R=\log_2\frac{2(\tau_A+\tau_B)\mu^{\xi-1}}{e|\tau_A-\tau_B|\chi^{\xi}}+h(\nu)-h\left(\frac{\sqrt{\lambda\lambda'}}{|\tau_A-\tau_B|}\right)\label{aa2}.
\end{aligned}
\end{equation}

Follow the discussion in Ref. 49, two situations need to be considered: one is that Alice and Bob know the transmissivities ($\tau_A$ and $\tau_B$) and the thermal noise affecting each link ($\omega_A$ and $\omega_B$). Then they must calculate the lower bound of the key rate for various $g$ and $g'$. To determine Eve's best strategy and give a lower bound of the key rate is of central importance in any QKD protocol~\cite{Gisin:2002:Quantum,Scarani:2009:The,Weedbrook:2012:Gaussian}, since overestimate of the rate will harm the security of the final key.
The minimized key rate of Eq.~\eqref{aa3} and Eq.~\eqref{aa2} in this situation can be written as
\begin{equation}
\begin{aligned}\label{eq1}
R(\tau,\omega_A,\omega_B)=&h\left(\frac{\tau+\lambda_{opt}}{\tau}\right)
+\log_2\frac{8\tau\mu^{\xi-1}}{e^2\chi_{opt}^{\xi}\lambda_{opt}},
\end{aligned}
\end{equation}
and
\begin{equation}
\begin{aligned}\label{eq2}
R(\tau_A,\tau_B,\omega_A,\omega_B)=&h\left(\frac{\tau_A+\lambda_{opt}}{\tau_B}\right)-h\left(\frac{\lambda_{opt}}{|\tau_A-\tau_B|}\right)+\log_2\frac{2(\tau_A \tau_B)^{\xi}\mu^{\xi-1}}{e|\tau_A-\tau_B|(\tau_A+\tau_B+\lambda_{opt})^{\xi}},\\
\end{aligned}
\end{equation}
respectively, with $\lambda_{opt}=\kappa+u|g|_{max}$ and $\chi_{opt}=2(2\tau+\lambda_{opt})/\tau$. The two expressions can be obtained by using the following two steps. First, by using the condition $g=-g'$, one can minimize the key rate. Second, by using the condition $g=|g|_{max}$, we can further minimize the key rate.

\begin{figure*}
\centering
\resizebox{12cm}{!}{\includegraphics{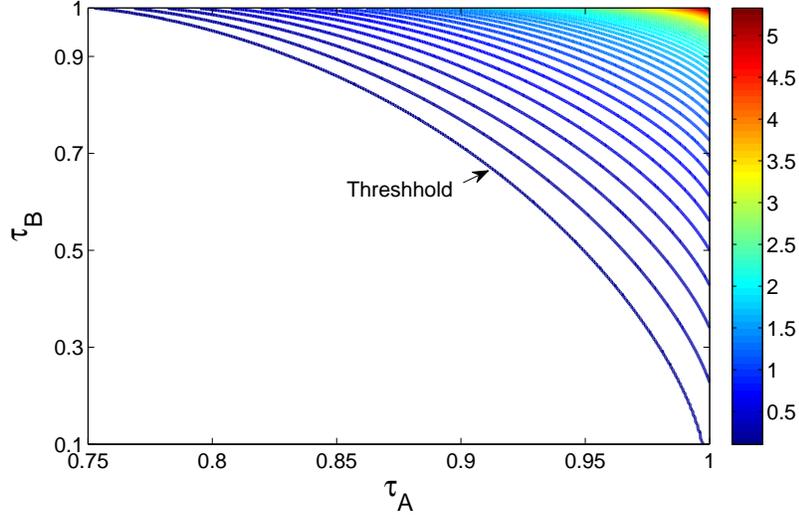}}
\caption{The secure key rate of CV-MDI-QKD varies with the transmissivities $\tau_{A}$ and $\tau_{B}$. Here, we use the parameters $\xi=0.97$, $\varphi=60$
and $\varepsilon=0.01$ for simulation. The unit of the secret key rate is bits per relay use.}
\label{fig:3}
\end{figure*}

The second situation is that Alice and Bob know the transmissivities and the equivalent noise ($\chi$). Actually, this is a more realistic situation since these parameters can be determined through data comparison in the classical post-process of the QKD protocol~\cite{pirandola:2015:high}. So the expression of the minimized key rate in symmetric condition is
\begin{equation}
\begin{aligned}\label{eq3}
R(\tau,\chi)=&h\left(\frac{\chi-2}{2}\right)+\log_2\frac{16\mu^{\xi-1}}{e^2\chi^{\xi}(\chi-4)},
\end{aligned}
\end{equation}
and in asymmetric condition is
\begin{equation}
\begin{aligned}\label{eq4}
R(\tau_A,\tau_B,\chi)=&\log_2\frac{2(\tau_A+\tau_B)\mu^{\xi-1}}{e|\tau_A-\tau_B|\chi^{\xi}}+h\left(\frac{\tau_A\chi}{\tau_A+\tau_B}-1\right)-h\left[\frac{\tau_A\tau_B\chi-(\tau_A+\tau_B)^2}{|\tau_A-\tau_B|(\tau_A+\tau_B)}\right],
\end{aligned}
\end{equation}
with $\chi=2(\tau_A+\tau_B)/\tau_A\tau_B+\varepsilon$, and $\varepsilon$ is the excess noise. The above two expressions can be achieved when $g=-g'$. So in the above two situations (thermal noise and equivalent noise), the condition $g=-g'$ can always minimize the key rate.
Based on Eq.~(\ref{eq4}) and the experiment parameters of Ref. 49, the secure key rate of CV-MDI-QKD for various transmissivities $\tau_{A}$ and $\tau_{B}$ is shown in Fig.~\ref{fig:3}.
We can see the secure key rate is smaller when the untrusted relay is closer to the center between Alice and Bob. The ideal situation is that the untrusted relay is very close to Alice or even the CV BSM is performed by Alice. Once Bob's raw key is used as the reference, the conclusion is opposite due to the symmetry of CV-MDI-QKD protocol.

We remark that Eq.~\eqref{aa3} to \eqref{eq4} are a natural generalization of the secure key rate formulas given in Ref. 49 from $\xi=1$ to a more general condition with $\xi\leq 1$. Here, we assume the condition that minimize the rate when $\xi=1$ also holds for $\xi<1$. In the next section, we will rigourously prove this assumption and shows the above generalization is valid.

\section*{Discussion}
In this paper, a new optical kind of CV-MDI-QKD scheme has been proposed. Through delicately manipulating the polarization, two laser pulses quantum states transmitted through the same fibers before CV BSM in the relay, thus their relative phase fluctuation is negligible and the phase-reference is self-aligned without introducing any complicated technologies.
Furthermore, we give an alternate proof of the minimized key rate condition to generate the secure key rate. We hope that our work can help the experimental study of the CV-MDI-QKD. One should note that the remote distance phase-locking technology may be achieved by the development of frequency combs or atom-clock synchronization.

We should point out that there are still some drawbacks in our scheme. Like all plug-and-play type QKD systems~\cite{Muller:1997:plug}, there exists untrusted source problem in our scheme~\cite{Zhao:2008:untrusted,Xu:2015:Measurement}. The most well-known threat for an untrusted source QKD protocol is the Trojan-horse attack~\cite{Jain:2014:Trojan,Lucamarini:2015:Practical}. Recently, a work shows Trojan-horse attack will greatly decrease the key rate along with the increasing of the mean photon number of the Trojan-horse mode~\cite{Jason:2018:hacking}. The practical security bound against the Trojan-horse attack in DV-QKD has been shown~\cite{Lucamarini:2015:Practical}. The security of collective attack in plug-and-play CV-QKD system has also been given~\cite{Huang:2016:continuous}. The wavelength filter and the intensity monitoring detector are the most suitable countermeasures for the Trojan-horse-type attack~\cite{Zhao:2008:untrusted,Xu:2015:Measurement,Lucamarini:2015:Practical,Huang:2016:continuous}. Note that these countermeasures will inevitably reduce the key rate. However, the unconditional security proof of CV-QKD with untrusted source is still an important open problem for future study.
Similar with the plug-and-play QKD system~\cite{Muller:1997:plug}, another drawback of our scheme is the strong Rayleigh scattering, which will effect the coherent detection at Charlie. This drawback has been studied in the work of the plug-and-play CV-QKD~\cite{Huang:2016:continuous}. Several methods have been presented to solve this problem, such as using a wideband shot-noise-limited
homodyne detector and preparing optical pulses with a narrow full width at half maximum~\cite{Huang:2016:continuous}. At last, the repetition rate is limited in the plug-and-play QKD system.

Surprisingly, one could use the new scheme to build a CV-MDI-QKD
network with a single untrusted source by further security analysis, which is similar with the DV-MDI-QKD network with an untrusted source~\cite{Xu:2015:Measurement}. Recently, the composable security against coherent attacks of CV-MDI-QKD has been proven~\cite{Lupo:2018:Continuous}.

\section*{Methods}
\textbf{The detailed proof of the minimized key rate condition.}
Due to the fact that $R$ is the same under the transformation $g\longleftrightarrow -g'$, $R$ is symmetric with respect to the bisector  $g=-g'$ \cite{pirandola:2015:high}.
It should to be noticed that the minimized key rate condition $g=-g'$ has been proven and used in Ref. 49. Here, we give an alternate proof of the minimized key rate condition by using the differential method. Our proof is quite straightforward. Under different situations, the key rate is a monotonic increasing function with the corresponding variable through the positivity of its first derivative. Then the minimum of $R$ is achieved once the variable reaches its minimum.

Consider any accessible point $(g_0,g'_0)$, the distance between this point and bisector $g=-g'$ is $d$ and $(l,-l)$ is the projection point of $(g_0,g'_0)$ on bisector $g=-g'$. With simple calculation, we have $d=|g_0+g'_0|/\sqrt{2}$ and $l=(g_0-g'_0)/2$. Due to the symmetry, we only need to consider the sector $g_0+g'_0\geq 0$, which leads to $g_0=d'+l$ and $g'_0=d'-l$ with $d=(g_0+g'_0)/\sqrt{2}$ and $d'=\sqrt{2}d/2$. Therefore, we have obtained the parameter transformation from $\{g,g'\}$ to $\{d',l\}$.

Once $\omega_A$ and $\omega_B$ are fixed ($\kappa$ is fixed), we let $\lambda =\kappa-u(d^{\prime}+l)$, $\lambda^{\prime}=\kappa +u(d^{\prime}-l)$,
we have $\lambda +\lambda ^{\prime }=2\delta$ and $\lambda \lambda ^{\prime }=\delta ^{2}-u^{2}d^{\prime 2}$
with $\delta =\kappa -ul>0$. The minimization procedure should be considered over two parameters $d'$ and $l$.
Once $\chi =\frac{\beta}{\alpha}\sqrt{(\beta+\lambda)(\beta+\lambda^{\prime})}$ is fixed,
the analytical expression of $\delta$ can be given by $\delta=\sqrt{\alpha^{2}\chi^{2}/\beta^{2}+u^{2}d^{\prime 2}}-\beta$  with $\alpha=\tau_A\tau_B$ and $\beta=\tau_A+\tau_B$. Meanwhile, we have $\lambda +\lambda ^{\prime } =2\left( \sqrt{\alpha ^{2}\chi ^{2}/\beta ^{2}+u^{2}d^{\prime 2}}-\beta \right)$
and $\lambda \lambda ^{\prime } =\alpha ^{2}\chi ^{2}/\beta ^{2}+\beta^{2}-2\beta \sqrt{\alpha ^{2}\chi ^{2}/\beta ^{2}+u^{2}d^{\prime 2}}$.
So if $\chi$ is fixed, we only need to minimize the key rate over the parameter $d^{\prime}$.

Here we show an alternate proof that $g=-g'$ minimize the key rate whenever the thermal noise or the equivalent noise is fixed for symmetric condition with $\tau_A= \tau_B$.
First, we consider the case of the fixed thermal noise. Under the symmetric condition, the expression of the key rate in Eq.~\eqref{aa3}  can be written as
\begin{equation}
\begin{aligned}\label{eq9}
R(y)=h(\nu_{1})-\log _{2}\nu_{2}-\log _{2}\nu_{3}+\log _{2}\frac{8}{e^{2}},
\end{aligned}
\end{equation}
where $\nu_{1}=\sqrt{(\tau+\delta )^{2}-y}/\tau$, $\nu_{2}=\mu^{1-\xi}\left(2\sqrt{(2\tau+\delta)^{2}-y}/\tau\right)^{\xi}$
and $\nu_{3}=\sqrt{\delta^{2}-y}/\tau$ with $y=u^{2}d^{\prime 2}$ and $0\leq y\leq \delta^2$. Obviously, $\nu_{3}<\nu_{1}$.

Next, we prove that $R(y)$ is monotonic increase function, so it reaches its minimum once $y=0$.
The first derivative of $R^{\prime}(y)$ is
\begin{equation}
\begin{aligned}\label{eq9}
R^{\prime }(y)=\frac{1}{2\tau^{2}}\left( \frac{\log _{2}e}{\nu _{3}^{2}}-\frac{1}{2\nu _{1}}g(\nu _{1})\right) -(\log _{2}\nu_{2})' >\frac{F(y)}{2\nu _{3}\tau^{2}}-(\log _{2}\nu_{2})',
\end{aligned}
\end{equation}
with $g(x)=\log_2\frac{x+1}{x-1}$, $F(y) =\frac{\log _{2}e}{\nu _{3}}-\frac{1}{2}g(\nu _{1})$ and $F^{\prime }(y) =\frac{\log _{2}e}{2\tau^{2}}\left[ \frac{1}{\nu _{3}^{3}}-\frac{1}{\nu _{1}(\nu _{1}^{2}-1)}\right]$. It's easy to verify that $g'(x)<0$ and $(\log _{2}\nu_{2})'<0$.
Due to the fact that $\nu _{1}^{2}-\nu _{3}^{2} =\frac{(\tau+\delta )^{2}-\delta ^{2}}{\tau^{2}}=\frac{\tau^{2}+2\tau\delta }{\tau^{2}}>1$,
we get $\nu _{1}^{2}-1>$ $\nu _{3}^{2}$, which leads to $F^{\prime}(y)>0$ and $F(y)\geq L(0)$.
Since $F(0)=F(0,\delta )=\frac{\tau\log _{2}e}{\delta }-\frac{1}{2}\log _{2}\frac{2\tau+\delta }{\delta }$, $F^{\prime }(0,\delta )=-\tau\log _{2}e\left[ \frac{1}{\delta ^{2}}-\frac{1}{\delta (\delta +2\tau)}\right] <0$ and together with the fact that $F(0,\delta \rightarrow +\infty )\rightarrow 0$, we have $F(y)\geq F(0,\delta )>0$. Then $R^{\prime}(y)>0$, so $R(y)$ reaches its minimum when $y=0$, which means $d^{\prime }=0$, i.e., $g=-g^{\prime }$.

The second situation is that the equivalent noise $\chi$ is fixed. Under this situation, the key rate of Eq.~\eqref{eq2} is simplified
under the symmetric condition as
\begin{equation}
\begin{aligned}
R=h(\nu_{1})-\log _{2}\nu_{2}+\log _{2}\frac{8\mu^{\xi-1}}{\chi^{\xi} e^{2} },
\end{aligned}
\end{equation}
where $\nu_{1}=\sqrt{(\tau+\lambda )(\tau+\lambda ^{\prime })}/\tau=\sqrt{b_{1}-a_{1}y}$, and $\nu_{2}=\sqrt{\lambda \lambda ^{\prime }}/\tau=\sqrt{b_{2}-a_{2}y}$,
with $y=\sqrt{\tau^{2}\chi^{2}/4+u^{2}d^{\prime 2}}$, $b_{1}=\frac{\chi ^{2}}{4}+1$, $a_{1}=\frac{2}{\tau}$,
$b_{2}=\frac{\chi ^{2}}{4}+4$ and $a_{2}=\frac{4}{\tau}$.
It is easy to check that $\nu_{1}>\nu_{2}$.

Next, we prove that $R(y)$ is a monotonic increasing function. The first derivative of $R(y)$ is \begin{equation}
\begin{aligned}
R^{\prime }(y)=\frac{a_{2}\log _{2}e}{2\nu _{2}^{2}}-\frac{a_{1}}{4\nu_{1}}g(\nu_1)>\frac{L(y)}{2\nu_{1}},
\end{aligned}
\end{equation}
with $L(y)=\frac{a_{2}\log _{2}e}{\nu_{2}}-\frac{a_{1}}{2}g(\nu_1)$ and $L^{\prime}(y)=\left[\frac{a_{2}^{2}}{2\nu_{2}^{3}}-\frac{a_{1}^{2}}{2\nu_{1}(\nu_{1}^{2}-1)}\right] \log _{2}e$.
Due to $\chi \geq \beta ^{2}/\alpha =4$, one has $\upsilon _{1}^{2}-\upsilon _{2}^{2}=(a_{2}-a_{1})y-(b_{2}-b_{1})=\frac{2}{\tau}y-3\geq \frac{2}{\tau}\frac{\tau\chi }{2}-3=\chi-3\geq 1$. So
$\nu_{1}^{2}-1\geq \nu_{2}^{2}$ and along with the fact that $a_{2}>a_{1}$, we have
$L^{\prime}(y) >0$ which further indicates $L(y) \geq L(\frac{\tau\chi}{2})$. $L^{\prime}(\frac{\tau\chi }{2})=-\frac{4\log _{2}e}{\tau}\left[ \frac{1}{(\chi-4)^{2}}-\frac{1}{\chi (\chi -4)}\right] <0$ with $\chi$ being a variable, together with the fact that $L(\frac{\tau\chi }{2})\rightarrow 0$ when $\chi \rightarrow +\infty $, we have $L(\frac{\tau\chi }{2})>0$. Finally, $R^{\prime}(y)>L(y)/2\nu_1\geq L(\frac{\tau\chi }{2})/2\nu_1>0$.
So $R(y)$ is a monotonic increasing function with its minimum obtained once $y$ reaches its minimum. Due to the fact that $y$ reaches its minimum when $d^{\prime }=0$, then $g=-g^{\prime }$ minimize the key rate under the symmetric condition.

In the above discussion, we have provided an alternate proof that $g=-g'$ is Eve's best strategy in the symmetric condition no matter with the fixed thermal noise or equivalent noise. Next we analyze the general condition with $\tau_A\neq \tau_B$.

We first consider the situation with fixed equivalent noise. The key rate of Eq.~\eqref{eq4} can be expressed as
\begin{equation}
\begin{aligned}
R=h(\nu _{1})-h(\nu _{2})+\log_2\frac{2(\tau_A+\tau_B)\mu^{\xi-1}}{e|\tau_A-\tau_B|\chi^{\xi}},
\end{aligned}
\end{equation}
where $\nu _{1}=\sqrt{b_{1}-a_{1}y}$ and $\nu _{2}=\sqrt{b_{2}-a_{2}y}$ with
$y=\sqrt{u^{2}d^{\prime 2}+\alpha ^{2}\chi ^{2}/\beta ^{2}}$, $b_{1}=1+\tau_A^{2}\chi ^{2}/\beta ^{2}$, $a_{1}=\frac{2}{\tau_B}$,
$b_{2}=(\beta ^{2}+\alpha ^{2}\chi ^{2}/\beta ^{2})/(\beta^{2}-4\alpha )$ and $a_{2}=2\beta/(\beta ^{2}-4\alpha)$.
The domain of definition for variable $y$ is $[y_{min},y_{max}]=[\alpha\chi/\beta,(\alpha^2\chi^2/\beta^2+\beta^2)/2\beta]$.

Next, we prove that $R(y)$ is a monotonic increasing function with $y$. Then its minimum can be achieved
once $y$ reaches its minimum ($d'=0$ or equivalently $g=-g'$ ).
The first derivative of $R(y)$ is
\begin{equation}
\begin{aligned}
R^{\prime }(y)=\frac{a_{2}}{4\nu _{2}}g(\nu _{2})-\frac{a_{1}}{4\nu _{1}}g(\nu _{1}).
\end{aligned}
\end{equation}
Introducing the variable transformations $q(y)=\frac{\nu _{1}+\nu _{2}}{2}$ and $p(y)=\frac{\nu _{1}-\nu _{2}}{2}$,
rewrite the expression of $R^{\prime}(y)$ as
\begin{equation}
\begin{aligned}
R^{\prime }(y)=\frac{p^{\prime }(y)}{2}\left[ g(\nu _{1})+g(\nu _{2})\right] +\frac{%
q^{\prime }(y)}{2}\left[ g(\nu _{1})-g(\nu _{2})\right],\label{aa1}
\end{aligned}
\end{equation}
with $p^{\prime }(y)=\frac{a_{2}\nu _{1}-a_{1}\nu _{2}}{4\nu _{1}\nu _{2}}$ and $q^{\prime }(y)=-\frac{a_{2}\nu _{1}+a_{1}\nu _{2}}{4\nu _{1}\nu _{2}}<0$.
The relationship between $\nu_{1}$ and $\nu_{2}$ can be summarized as
\begin{equation}
\begin{aligned}
\nu _{1}>\nu _{2}\text{ ~ }(\tau_A\geq 2\tau_B), \\
\nu _{1}<\nu _{2}\text{ ~ }(\tau_A<2\tau_B).
\end{aligned}
\end{equation}
While there is an extra restriction for $\chi$ when $\nu _{1}<\nu _{2}$
\begin{equation}
\begin{aligned}
\chi\geq\frac{2\tau_B\beta }{\tau_A(2\tau_B-\tau_A)}\text{ ~ }(\tau_B<\tau_A\leq 2\tau_B),\\
\chi\geq \frac{2\beta }{\tau_A}\text{ ~ }(\tau_A<\tau_B),\label{rela_1}
\end{aligned}
\end{equation}
otherwise $\nu_{1}>\nu_{2}$.
Meanwhile, $p^{\prime}(y)>0$ holds for any value of the parameters.
For $\nu_1>\nu_2$, since $g(x)$ is a monotonically decreasing function, we have $g(\nu_1)<g(\nu_2)$. Alongside with
the fact that $q^{\prime}<0$ and $p^{\prime}>0$, it's easy to see that $R^{\prime}(y)>0$.

The remaining problem is to prove $R^{\prime}(y)>0$ also holds for $\nu_1<\nu_2$.
Introducing a new function $D(y)=d(\nu _{1})-d(\nu _{2})$ with $d(x)=(x-1)g(x)$. The first and second derivatives of $d(x)$ are
$d^{\prime }(x)=g(x)-\frac{2\log _{2}e}{x+1}$, and $d^{\prime \prime }(x)=\frac{2\log _{2}e}{x+1}(\frac{1}{x+1}-\frac{1}{x-1})<0$, respectively.
Since $d^{\prime}(x)\rightarrow 0\text{ }(x\rightarrow +\infty )$, we have $d^{\prime }(x)>0$. So $d(x)$ is a monotonically increasing function. Due to the fact that $\nu_{1}<\nu _{2}$,
i.e., $D(y)<0$, the following inequation holds $\frac{g(\nu _{1})}{g(\nu _{2})}<\frac{\nu _{2}-1}{\nu _{1}-1}$.
Thus, we have
\begin{equation}
\begin{aligned}
R^{\prime }(y)=\frac{a_{2}}{4\nu_{2}}g(\nu_{2})-\frac{a_{1}}{4\nu_{1}}g(\nu_{1}) >\frac{a_{1}}{4\nu_{2}}g(\nu_{2})\frac{A(y)}{a_{1}\nu_{1}(\nu_{1}-1)},
\end{aligned}
\end{equation}
where $A(y)=(a_{2}\nu_{1}^{2}-a_{1}\nu_{2}^{2})-(a_{2}\nu_{1}-a_{1}\nu_{2})
=(a_{2}\nu_{1}^{2}-a_{1}\nu_{2}^{2})-k(y)\geq (a_{2}\nu_{1}^{2}-a_{1}\nu_{2}^{2})-k(y_{\min})$
with $k(y)=a_{2}\nu_{1}-a_{1}\nu_{2}$.
Here we use the fact that $k(y)\leq k(y_{\min})$ when $\nu_{1}<\nu _{2}$ .
For $\tau_B<\tau_A\leq2\tau_B$,
\begin{equation}
\begin{aligned}
k(y_{\min })=\frac{4\tau_A\tau_B\chi }{(\tau_A+\tau_B)(\tau_A-\tau_B)^{2}}-\frac{2(\tau_A+\tau_B)(2\tau_B-\tau_A)}{\tau_B(\tau_A-\tau_B)^{2}},
\end{aligned}
\end{equation}
which leads to
\begin{equation}
\begin{aligned}
A(y)\geq \frac{2\tau_A^{3}\left[ \chi ^{2}-2\tau_B(\tau_A+\tau_B)\chi /\tau_A^{2}\right] }{(\tau_A^{2}-\tau_B^{2})^{2}}-\frac{2(\tau_A+\tau_B)}{\tau_B(\tau_A-\tau_B)}.
\end{aligned}
\end{equation}
Due to $\chi \geq \frac{2\tau_B(\tau_A+\tau_B)}{\tau_A(2\tau_B-\tau_A)}>\frac{\tau_B(\tau_A+\tau_B)}{\tau_A^{2}}$, we have
$A(y)\geq \frac{4\tau_A^{2}(3\tau_B-\tau_A)}{\tau_B(\tau_A-\tau_B)(2\tau_B-\tau_A)^{2}}>0$,
thus $R^{\prime }(y)>0$.
For $\tau_A<\tau_B$,
\begin{equation}
\begin{aligned}
k(y_{\min })=\frac{4\tau_A^{2}\chi }{(\tau_A+\tau_B)(\tau_A-\tau_B)^{2}}-\frac{2\tau_A(\tau_A+\tau_B)}{\tau_B(\tau_A-\tau_B)^{2}},
\end{aligned}
\end{equation}
which leads to
\begin{equation}
\begin{aligned}
A(y)\geq \frac{2\tau_A^{3}\tau_B\left[ \chi ^{2}-2(\tau_A+\tau_B)\chi /\tau_A\right] }{\tau_B(\tau_A^{2}-\tau_B^{2})^{2}}.
\end{aligned}
\end{equation}
Due to $\chi \geq \frac{2(\tau_A+\tau_B)}{\tau_A}>\frac{\tau_A+\tau_B}{\tau_A}$, we have $A(y)\geq 0$, thus $R^{\prime}(y)>0$.
Therefore, $R^{\prime}(y)>0$ is a general result independent of the values of parameters. In conclusion, $g=-g^{\prime}$ always minimized the key rate when $\chi$ is fixed.
As for the condition with thermal noise $\omega_A$ and $\omega_B$ fixed, we can follow the above procedures to prove $g=-g^{\prime}$ minimize the rate. For simplicity, we will not show the detailed calculations here. Follow the above discussion, in this situation the rate can be further minimized when $\lambda=\lambda_{opt}$, we also prove this conclusion.

\noindent
\textbf{The relationship between $\nu_1$ and $\nu_2$.}
Now we compare $\nu_{1}$ and $\nu_{2}$ through determining the sign of function
\begin{equation}
\begin{aligned}
\nu _{1}-\nu _{2}=\frac{(b_{1}-b_{2})-(a_{1}-a_{2})y}{\upsilon_{1}+\upsilon _{2}}=\frac{a_{1}-a_{2}}{\upsilon _{1}+\nu _{2}}\left( y_0-y\right),
\end{aligned}
\end{equation}
with $y_0=\frac{b_{1}-b_{2}}{a_{1}-a_{2}}$.
The above equation holds when $a_{1}\neq a_{2}$.

For $a_{1}=a_{2}$ ($\tau_A=3\tau_B$),
$\nu _{1}-\nu _{2}=\frac{b_{1}-b_{2}}{\upsilon_{1}+\upsilon _{2}}$.
Based on the fact that $\chi>\frac{2\beta }{\tau_A}$, it's easy to verify $b_{1}>b_{2}$ and $\nu _{1}>\nu _{2}$.

For $a_{1}<a_{2}$ ($\tau_A<3\tau_B$),$y\geq y_0$ means $\nu_{1} \geq\nu _{2}$ and $y<y_0$ means $\nu _{1} <\nu _{2}$.
Then, $y_{\max }-y_0>0$, here we use the fact that $\chi\geq\beta^2/\alpha$.
If $3\tau_B>\tau_A>2\tau_B$, and based on the fact that $\frac{\tau_A\chi }{\beta }>\frac{\beta }{\tau_B}>2$,
we have $y_{\min}> y_0$ which means $\nu_{1}>\nu_{2}$.
If $\tau_A<2\tau_B$, $y_{\min}\leq y_0$ holds only if $\chi \geq \frac{\beta (3\tau_B-\tau_A+|\tau_A-\tau_B|)}{\tau_A(2\tau_B-\tau_A)}$.
For $\tau_B<\tau_A<2\tau_B$, we have $\chi \geq \frac{2\tau_B\beta }{\tau_A(2\tau_B-\tau_A)}$.
For $\tau_A<\tau_B$, we have $\chi \geq \frac{2\beta }{\tau_A}$.
So with the proper value of $\chi$ , $\nu_{1}\leq \nu_{2}$ can be reached when $\tau_A<2\tau_B$.

For $a_{1}>a_{2}$ ($\tau_A>3\tau_B$), $y\leq y_0$ leads to $\nu_{1} \geq \nu_{2}$ while $y>y_0$ leads to $\nu_{1}<\nu_{2}$.
Following the same procedure, $y_{\max }-y_0<0$, so we have $\nu_{1}>\nu_{2}$. In a word, once $\tau_A\geq 2\tau_B$, we have $\nu_{1}>\nu_{2}$.

\noindent
\textbf{The Proof of $p^{\prime}(y)>0$.} The expression of $p^{\prime}(y)$ is $p^{\prime}(y)=\frac{a_{2}\nu_{1}-a_{1}\nu_{2}}{4\nu_{1}\nu_{2}}$.
Here we introduce a new function $k(y)=a_{2}\nu_{1}-a_{1}\nu_{2}$,
$k^{\prime}(y)=\frac{a_{1}a_{2}}{2}(\frac{1}{\nu_{2}}-\frac{1}{\nu_{1}})$.

For $\tau_A\geq 2\tau_B$, we have $\nu_{1}>\nu_{2}$. Then $k^{\prime }(y)>0$, so $k(y) \geq k(y_{\min})>0$, we have $p^{\prime}(y)>0$.
Next, we consider the condition $\tau_A<2\tau_B$. Based on Eq.~(\ref{rela_1}), once $\chi<\frac{\beta (3\tau_B-\tau_A+|\tau_A-\tau_B|)}{\tau_A(2\tau_B-\tau_A)}$, $\nu_{1}>\nu_{2}$ so we get $p^{\prime}(y)>0$. Once $\chi\geq\frac{\beta (3\tau_B-\tau_A+|\tau_A-\tau_B|)}{\tau_A(2\tau_B-\tau_A)}$, $y_{\min}\leq y_{0}<y_{max}$. The region of $y$ can be divided
into three parts
\begin{equation}
\begin{aligned}
y_{\min} &<y<y_{0}\text{, }k^{\prime}(y)<0\text{ }(\nu_{1}<\upsilon_{2}), \\
y &=y_{0}\text{, }k^{\prime }(y)=0\text{ }(\nu_{1}=\nu_{2}=\nu), \\
y_{0} &<y<y_{\max}\text{, }k^{\prime}(y)>0\text{ }(\nu_{1}>\nu_{2}).
\end{aligned}
\end{equation}
It's easy to see that $k(y)$ reaches its minimum when $\nu_{1}=\nu_{2}$, so $k(y)=a_{2}\nu_{1}-a_{1}\nu_{2}\geq (a_{2}-a_{1})\nu>0$,
we have $p^\prime(y)>0$. Therefore, $p^\prime(y)>0$ is always satisfied independent of the values of parameters.

\noindent
\textbf{Minimization over $\lambda$.}
Once the thermal noise is fixed, the minimization is actually a two-step procedure. We have already proved $g=-g'$ minimize the key rate, next is to prove
$\lambda=\lambda_{opt}$ further minimize the key rate.
The key rate can be divided into the following two parts
\begin{equation}
\begin{aligned}
R=H(\tau_A,\tau_B,\lambda )+L(\tau_A,\tau_B,\lambda),
\end{aligned}
\end{equation}
with $H(\tau_A,\tau_B,\lambda )=h(\frac{\tau_A+\lambda }{\tau_B})-h(\frac{\lambda }{|\tau_A-\tau_B|})$, and
$L(\tau_A,\tau_B,\lambda )=\log_2\frac{2(\tau_A \tau_B)^{\xi}\mu^{\xi-1}}{e|\tau_A-\tau_B|(\tau_A+\tau_B+\lambda)^{\xi}}$.

It is easy to see that $L(\tau_A,\tau_B,\lambda)$ is minimized when maximizing $\lambda $, so the remaining part is to prove that $H(\tau_A,\tau_B,\lambda )$ is
also minimized under this same condition. To find the minimum of $H(\tau_A,\tau_B,\lambda)$, we give its first and second derivatives
$H^{\prime }(\tau_A,\tau_B,\lambda )=\frac{1}{2\tau_B}g(\frac{\tau_A+\lambda }{\tau_B})-\frac{1}{2|\tau_A-\tau_B|}g(\frac{\lambda }{|\tau_A-\tau_B|})$,
and $H^{^{\prime \prime }}(\tau_A,\tau_B,\lambda )=\frac{1}{\lambda^{2}-|\tau_A-\tau_B|^{2}}-\frac{1}{(\lambda +\tau_A)^{2}-\tau_B{}^{2}}$, respectively. Then the sign of $H^{^{\prime \prime }}(\tau_A,\tau_B,\lambda)$ is determined by the sign of the following term
$(\lambda ^{2}-|\tau_A-\tau_B|^{2})-[(\lambda +\tau_A)^{2}-\tau_B{}^{2}]=-2\tau_A(\lambda +\tau_A-\tau_B)\leq -2\tau_A(\lambda -|\tau_A-\tau_B|)<0$.
Here, we use the fact that $\lambda >|\tau_A-\tau_B|$. If $\lambda$ satisfies $\lambda \leq |\tau_A-\tau_B|$, a clear consequence is that $\frac{\lambda}{|\tau_A-\tau_B|}\leq 1$
is existing. So $h(\frac{\lambda }{|\tau_A-\tau_B|})$ will give complex value which is certainly non-physical. So to overcome this problem, we have to set $\lambda >|\tau_A-\tau_B|$,
and this further confirms $\frac{\tau_A+\lambda }{\tau_B}>1$ to make the term $h(\frac{\tau_A+\lambda }{\tau_B})$ physical. So $H^{^{\prime \prime }}(\tau_A,\tau_B,\lambda )> 0$,
makes $H^{\prime}(\tau_A,\tau_B,\lambda )$ monotonically increasing in the region. For $\lambda\rightarrow +\infty $, $H^{\prime }(\tau_A,\tau_B,\lambda )\rightarrow 0$, combined with the fact that it increases monotonically, we obtain $H^{\prime}(\tau_A,\tau_B,\lambda )<0$. So $H(\tau_A,\tau_B,\lambda)$ monotonically decreases in the region, and it reaches its minimum when maximizing $\lambda$, which
equals $\lambda_{opt}=\lambda+u|g|_{max}$. Based on the deduction above, we prove that $R$ reaches its minimum when $\lambda$ reaches its maximum.
For simplicity, we only give the proof for asymmetric condition above. It's easy to prove the conclusion also holds for the symmetric condition.

\noindent\textbf{Acknowledgments}\\
H.-L. Yin gratefully acknowledges support from the Fundamental Research Funds for the Central Universities.

\noindent\textbf{Author Contributions}\\
H.-L.Y. and Y.F. have the main idea. All results are acquired through the
discussion among all authors. All authors contribute to the writing and reviewing of the manuscript.

\noindent\textbf{Additional Information}\\
Competing interests: The authors declare no competing interests.

\bibliographystyle{naturemag}


\end{document}